\documentclass[12pt,preprint]{aastex}

\newcommand{\NHI}{$N($\ion{H}{1}$)$}
\newcommand{\icm}{cm$^{-2}$}
\newcommand{\hkpc}{$h^{-1}$ kpc}
\newcommand{\he}{HE0512$-$3329}

\newcommand{\kms}{km~s$^{-1}$}

\newcommand{\lya}{Ly$\alpha$\ }
\newcommand{\lyb}{Ly$\beta$\ }

\slugcomment{Revised version, Feb 22 2005}

\shorttitle{Damped \lya System in Two Lines of Sight}
\shortauthors{Lopez et al.}

\begin{document}


\title{Metal Abundances in a Damped \lya System
  Along Two  Lines of Sight at $z=0.93$\altaffilmark{1} }       

\altaffiltext{1}{Based on observations with the NASA/ESA {\it
Hubble Space Telescope}, obtained at the Space Telescope Science
Institute, which is operated by the Association of Universities for
Research in Astronomy, Inc., under NASA contract NAS 5-26555.
Also based on data obtained at the ESO Very Large
Telescopes, program  70.A-0446(A).
}

\author{Sebastian Lopez\altaffilmark{2}, 
Dieter Reimers\altaffilmark{3},
Michael D. Gregg\altaffilmark{4,7}, 
Lutz Wisotzki\altaffilmark{5}, 
Olaf Wucknitz\altaffilmark{6}, 
and Andres Guzman\altaffilmark{2}
} 

\altaffiltext{2}{
Departamento de Astronom\'{\i}a, Universidad de Chile, Casilla 36-D,
Santiago, Chile. }

\altaffiltext{3}{
Hamburger Sternwarte, Gojenbergsweg 112, D-21029 Hamburg, Germany
}
\altaffiltext{4}{
Physics Department, University of California, Davis,
CA 95616
}
\altaffiltext{5}{
Astrophysikalisches Institut Potsdam,
An der Sternwarte 16, D-14482 Potsdam, Germany
}
\altaffiltext{6}{
Institut f\"ur Physik, Universit\"at Potsdam, Am Neuen Palais,
D-14469 Potsdam, Germany
}
\altaffiltext{7}{Institute of Geophysics \& Planetary Physics,
Lawrence Livermore National Laboratory, Livermore, CA 94550}

\begin{abstract}

We study metal abundances in the $z=0.9313$ damped \lya system
observed in the two lines-of-sight, A and B, toward the
gravitationally-lensed double QSO \he.  Spatially resolved Space
Telescope Imaging Spectrograph spectra constrain the neutral-gas
column density to be $N($\ion{H}{1}) $= 10^{20.5}$ \icm\ in both A
and B.  UV-Visual Echelle Spectrograph spectra (spectral resolution
FWHM $=9.8$ \kms) show, in contrast, significant line-of-sight
differences in the column densities of \ion{Mn}{2} and \ion{Fe}{2};
these are not due to observational systematics.  We find that [Mn/H]
$=-1.44$ and [Fe/H] $=-1.52$ in damped \lya system A, while [Mn/H]
$-0.98$ and [Fe/H] $>-1.32$, and possibly as high as [Fe/H] $\approx
-1$ in damped \lya system B.  A careful assessment of possible
systematic errors leads us to conclude that these transverse
differences are significant at a $5\sigma$ level or greater.  Although
nucleosynthesis effects may also be at play, we favor differential
dust-depletion as the main mechanism producing the observed abundance
gradient.  The transverse separation is 5 $h^{-1}_{70} $kpc at the
redshift of the absorber, which is also likely to be the lensing
galaxy.  The derived abundances therefore probe two opposite sides of
a single galaxy hosting both damped \lya systems.  This is the first
time firm abundance constraints have been obtained for a single damped
system probed by two lines-of-sight.  The significance of this finding
for the cosmic evolution of metals is discussed.
\end{abstract}

\keywords{cosmology: observations --- galaxies: abundances ---
intergalactic medium --- quasars: absorption lines --- quasars:
individual (\he)}

\section{Introduction}

Measuring metallicities with $10\%$ precision in high$-z$ damped \lya systems
(e.g., Prochaska \& Wolfe 2002) has become a standard technique for tracing
chemical evolution in neutral gas over a large fraction of a Hubble time.
Surveys of damped \lya systems (DLAs) contrast observations with models of
chemical evolution, attempting to understand the nature of the objects hosting
the absorbing clouds and to assess the degree of metal enrichment of gas in
high redshift galaxies.  The anticipated increase of [M/H] with cosmic time
has become apparent only recently with the dramatic increase in sample size
over the last few years (Prochaska et al.\ 2003a).  However, this evolutionary
trend is hidden by a considerable scatter in [M/H] of $\approx 1$
dex\footnote{In the usual convention of [M/H] $\equiv \log({\rm
M/H})-\log({\rm M/H})_{\sun}$}, that is seen at every epoch from $z=5$ to
$z=0$.  The observed cosmic abundance spread at a given epoch is attributed to
two distinct effects: (1) lines of sight to background QSOs probe a single
location in DLAs at a random impact parameter, so gradients in DLAs will
produce different abundances and dust-to-gas ratios, and (2) DLAs arise in a
variety of galaxy types with different star formation and enrichment
histories, as observed at low redshift (e.g., Boisse et al.\ 1998; Rao et al.\
2003).  Adding to the scatter, the overall census of DLAs might be biased
toward lower metallicities because QSOs in front of very dusty and metal-rich
absorbers might be missed entirely by optical surveys (e.g., Pei, Fall \&
Bechtold 1991; but see Ellison et al.\ 2001).

DLAs probed by two or more lines-of-sight offer a way to test these
theories of the observed scatter in [M/H] while also further probing
the nature of DLAs.  The small impact parameters involved make
gravitationally lensed QSOs the ideal tool for studying DLAs this way.
The number of lens systems at high redshift, however, is small and so
also the number of intervening DLAs (Le Brun et al. 2000; Smette,
Claeskens, \& Surdej 1997).  The relative faintness of QSOs and the
small separations on the sky make spatially resolved, high-resolution
spectroscopy challenging, and from the few cases studied, no separate
DLA abundances have been obtained (Smette et al.\ 1995; Zuo et al
1997; Lopez et al.\ 1999; Kobayashi et al. 2002; Churchill et al.\
2003; Ellison et al.\ 2004a).

In the course of a Hubble Space Telescope imaging snapshot survey using STIS
to search for small separation gravitationally lensed quasars, \he\ was
discovered (Gregg et al.\ 2000).  This system is a double image QSO at
$z_{em}=1.58$ with a separation of just $0\farcs644$.  The extinction and
microlensing properties have been investigated by Wucknitz et al.\ (2003;
hereafter 'Paper I').  In this work, we focus on the metal abundances in the
DLA observed along the two lines of sight toward \he.  The position angle of
the two point sources is $17\arcdeg$ N to E, with the brighter component
(herafter 'A') to the North.  The small separation of \he\ makes ground-based
spectroscopy extremely challenging and since achieving cleanly separated
spectra is crucial for the present study, a good fraction of this paper is
devoted to a description of data reduction (\S~\ref{section_obs}) and possible
systematic effects (\S 3).  The derived abundances at $z=0.9313$ are presented
in \S\ 4, again with careful consideration of possible systematics.  The
results are discussed in \S 5, and a summary is given in \S 6.  We adopt a
$H_0=70$ \kms~Mpc$^{-1}$, $(\Omega_M,\Omega_\Lambda)=(0.3,0.7)$ cosmology.

\section{Observations and Data Reduction}
\label{section_obs}

\subsection{HST STIS observations}

Spatially resolved spectra of \he\ A and B were obtained with the
Space Telecope Imaging Spectrograph (STIS) in first-order grating
spectroscopic mode using G230L with the MAMA detector and G430L with
the CCD.  Details of the observations and data reduction are found in
Paper I.  The wavelength coverage is $1570$ to $5700$
\AA\ at a spectral resolution of $3.3$ \AA\ in the blue and $4.1$ \AA\
in the red, or $\sim 300$ \kms.  The signal-to-noise is S/N $\sim 30$ 
in all regions of interest.  This resolution and S/N are not
high enough to determine column densities of metal species, whose
absorption lines are only a few \kms\ wide, but it is quite suitable
for measuring the \ion{H}{1} column density from the $z=0.93$ damped
\lya line observed at $\lambda=2350$ \AA.  The $N($\ion{H}{1}) is needed
to measure the gas metallicity.

\subsection{VLT UVES observations and Data Reduction}
\label{section_vlt_obs}

\he\ was observed using the VLT UV-Visual Echelle Spectrograph (UVES)
on the nights of January 2 and 3 2003.  The spectra of both QSOs were
acquired simultaneously by aligning the slit with the 2 images.  This
was judged the best way to quantify the amount of cross-talk between
the partially resolved spectra.  Since Paranal median seeing values
are critically close to the separation between QSO images, service
mode was preferred (with its associated caveats, see below).  The
dichroic '$390+564$' setup was used, which covers $3600-4800$ \AA\ in
the blue arm, and $4600-7500$ \AA\ in the red one.  Eight $3000$
seconds exposures were obtained, with seeing monitor values ranging
from $0\farcs6$ to $0\farcs9$.  The slit width was 1\farcs0 and the
pixel scale is $0\farcs18$ in the red CCD, but only $0\farcs25$ in the
blue.  Under these circumstances, extraction of cleanly separated
spectra becomes challenging.

\begin{figure}
\epsscale{.40}
\plotone{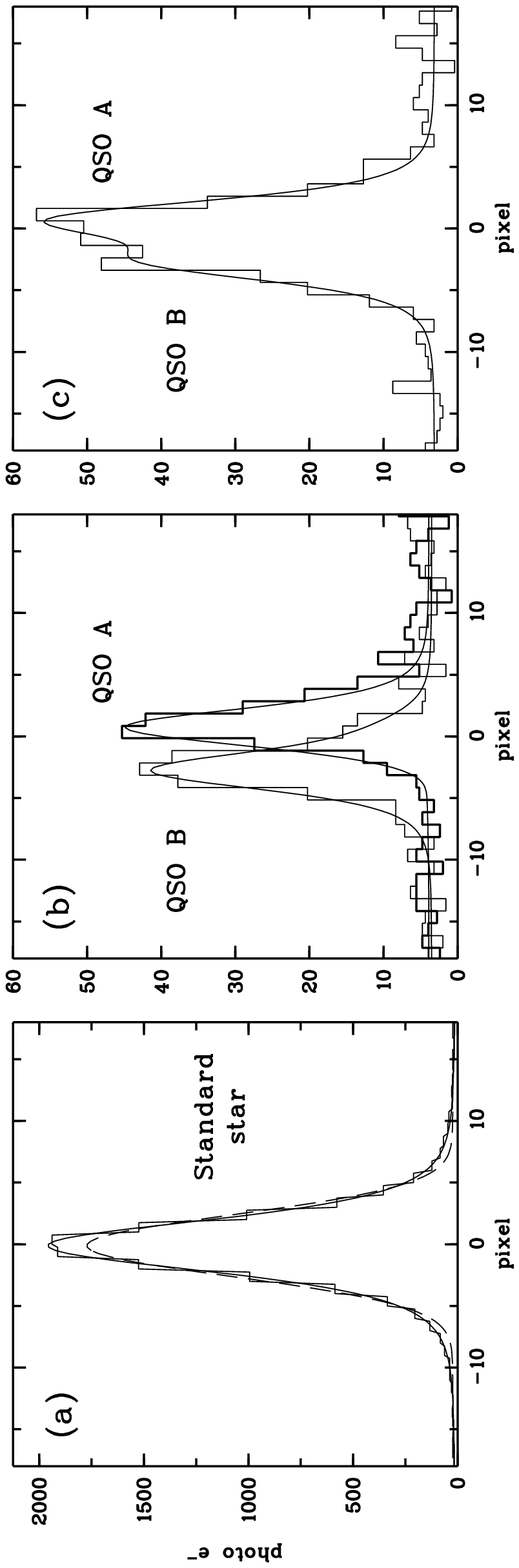}
\caption{
Spatial profiles in echelle orders of single exposures with UVES
grating 564 (pixel size, $0\farcs18$). The orders have been rectified
and binned $\times4$ in the dispersion direction. $(a)$ Standard star at
seeing of $0\farcs8$. The dashed line is a best-fit Gaussian; the
smooth line is a best-fit Moffat, taken to best represent the point-spread
function. $(b)$ \he\ A and B at two different wavelengths where black
aborption by \ion{Mg}{2} occurs in only one spectrum. Moffat
best-fit profiles are overlaid. (c) \he\ A
and B  at an unabsorbed wavelength. In $(a)$ and $(b)$ the seeing was
$0\farcs63$ and the separtion between QSO images is $0\farcs64$.    
\label{fig_reduction}}
\end{figure}

To obtain separate spectra of \he\ A and B, we used our own reduction
pipeline to extract spectra of multiple objects in echelle data. The
algorithm is based on a simultaneous PSF fit to each of the seeing
profiles with consideration of pixel variances. The fits are performed
twice: first allowing all PSF parameters to vary, then constraining
PSF width and position by modelling them with low-order polynomials
along the spectral direction. The extracted flux is defined by the
synthetic profiles, and a $1\sigma$ flux error array results from
propagation of errors (Lopez et al.\ 1999).  In \he, the spatial
profile was found to be well-represented by a Moffat function (Moffat
1969) with best-fit index parameter $\beta=2.5$ (see
Fig.~\ref{fig_reduction}a).

Besides spatial resolution, low S/N is also a problem in the present
UVES data.  
On-chip binning in the spectral direction only could not be carried out
because this was not offered in service mode. 
Attempts to extract the
QSO spectra from the unbinned data were unsuccessful, resulting in
large flux residuals, leading us to bin the data by 4 pixels in the
dispersion direction; the data have an initial FWHM=$\sim 6$ pixels,
so there is margin enough for rebinning.  To this end, the echelle
orders were first 'unbent'.  Fortunately, the use of gratings as
cross-dispersers makes UVES data free from distortions in the
spatial direction (no emission line 'tilts'), so the order
rectification is not critical.  Rectification was done with our 
own  script,  which traces the echelle orders with a Hough
transform (Ballester 1994), re-samples the data in the spatial
direction, computes the distance (in fractions of a pixel) to the
center of the order definition, and re-allocates pixels with respect to
a new, straight order definition.  
The rectified orders could then be
re-binned $\times4$ along the spectral direction ---now aligned with the CCD
rows--- thus increasing the S/N without 
distorting the spatial profile. The goodness of the treatment is
illustrated in Fig.~\ref{fig_reduction}a, where the spatial profile of
the standard star keeps its symmetry and Moffat-function-like form
after the two processes.

Even at the improved S/N, the closeness of the two objects did not
allow us to robustly determine a PSF width from global fits along the
orders.  Instead, this parameter (the actual seeing) was obtained from
single-profile fits at wavelengths where one of the QSOs is completely
absorbed by saturated $z=0.9313$ \ion{Mg}{2}
(Fig.~\ref{fig_reduction}b).  The spatial separation between QSO
profiles was held fixed at $0\farcs644$, as determined from the
space-based observations.

We found that the extracted spectrum of QSO A does not reach zero at the
position of \ion{Mg}{2} where we expect totally black absorption.  One
possible explanation for this residual is emission by the lensing galaxy at
$z=0.93$.  To explore this, we averaged columns all the way along the
absorption troughs -- some tens of pixels -- to obtain an averaged QSO profile
(the rectified orders permit this operation).  Up to 10\% flux excess is
present to the North of {\it both} QSOs.  This PSF asymmetry is barely
apparent in Fig.~\ref{fig_reduction}b [although the Figure shows just one
column], and, although it explains the residual flux in QSO A, it seems to
exclude the possibility that the excess is from the lensing galaxy  (this
is because any emission excess between the two QSO images would appear to the
North of QSO B only).  The extracted spectra were corrected by subtracting a
pedestal at the level of up to 10\% of the QSO B flux.  The possible effect of
this correction on absorption line parameters will be considered below.

Finally, to wavelength calibrate the extracted orders, exactly the
same pre-processing was applied to the Th-Ar exposures.  The final
wavelength solution is accurate to rms $\sim 1/7$ pixel.  Before
coadding orders, the spectra were rebinned to a common
vacuum-heliocentric scale with pixel size of $0.0994$ \AA.  The final
spectral resolution in the coadded spectra is FWHM $\sim 9.7$ \kms\
($\sim 2$ binned pixels), and the final S/N per pixel is $\approx 35-40$ for
both A and B.  Finally, the coadded orders were normalized directly by
simultaneously dividing out the response function and QSO continuum.
The continuum was estimated independently in A and B by fitting cubic
splines though featureless spectral regions.

Unfortunately, a proper extraction of the blue-arm spectra was
impossible given both the too poor spatial sampling and the lack of
absorbed regions, needed to characterize the PSF.  The final, usable,
wavelength coverage is $4620$ to $5600$ \AA.

\section{Column Densities and Assessment of Systematic Errors}

We used two methods to obtain column densities: the apparent optical
depth method (AODM; Savage \& Sembach 1991) and Voigt profile
fitting.  The fits were performed with the MIDAS package FITLYMAN
(Fontana \& Ballester 1995).  While both methods offer similar results
for weak transitions, strong blending is more realiably asssessed by
profile fitting.  Only metal lines believed to be non-saturated were
fitted (see below).  Oscillator strengths, $f$, were taken from the new
compilation by Morton (2000), and solar abundances from Grevesse \&
Sauval (1998).

While the UV STIS spectra are necessary for determining \NHI, accurate
determination of the metal abundances requires the much higher
spectral resolution provided by UVES.  These two datasets suffer
respectively from two possible systematic effects: bad continuum
estimate, and cross-talk between UVES spectra.  Since the main goal of
the present study is to detect possible line-of-sight differences in
the absorption line parameters, in the following we describe in detail
these sources of error.

\begin{figure}
\epsscale{.80}
\plotone{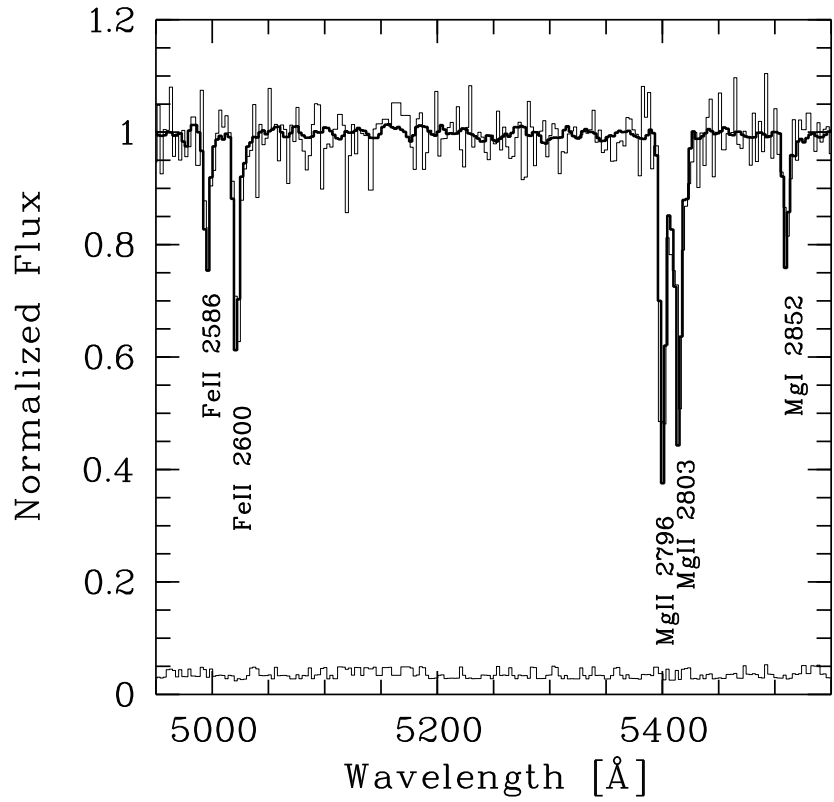}
\caption{
Comparison between STIS and UVES spectra of \he\ A redwards of \lya
emission. The UVES spectrum (thick line) was 
smoothed and rebinned to match the STIS specifications.  
Also shown is the $1\sigma$-error array of the STIS
flux. 
\label{fig_smooth}}
\end{figure}

As a zeroth order approach we compare the reduced UVES spectra with the STIS
data, which we assume to be fully resolved spatially.  In
Fig.~\ref{fig_smooth} we show a comparison between the spectra redwards of the
quasar \lya emission; the high resolution data have been smoothed and rebinned
to STIS specifications.  A $5$ \AA\ shift was applied to the HST spectra
to match the UVES vacuum-heliocentric wavelength scale (in the STIS
acquisition images we found an offset from the image center of about 2 pixels,
which, given the wide slit used, would correspond to a shift in the dispersion
direction of $\sim 5$ \AA). We note a good match in the
displayed transitions, which supports a proper extraction of the UVES
spectra.

Most of the lines detected in
the STIS spectra are actually saturated, and so this makes it impossible to
fairly compare column densities of A and B. Indeed, unsaturated transitions
are lost in the noise of the STIS data.  We attempted a comparison of
equivalent widths between (unconvolved) UVES and STIS non-saturated lines;
while consistency is found in a few cases, most of these lines are not detected
at the $3\sigma$ significance level in the STIS spectra. Thus, a 
comparison of STIS and UVES data serves as a valuable consistency check
between the ground and space-based observations; however, even flux
discrepancies in the UVES data as high as 10\%, or equivalent width
differences as high as $30$\% in {non-saturated} lines would be impossible to
detect at the modest S/N and spectral resolution of the STIS data.


To assess the impact of possible cross-talk effects on column
densities, we computed equivalent widths of several {\it weak} lines
in the UVES spectra, typically of $\sim 30$ m\AA, observed.  Mock
echelle orders  were then created in which 10\% of the --non-normalized-- flux
of one QSO  was
respectively added to and subtracted from the other QSO.  The mock
spectra were renormalized, and equivalent widths of the same
absorption lines were re-computed.  We found typical deviations of the
order $\sim 8$\% in equivalent width.  For intrinsically strong
transitions, e.g., \ion{Fe}{2} $\lambda 2586$, this equivalent width
corresponds to $N\approx 5\times 10^{12}$ \icm, and the differences
translate into 0.04 dex deviations.  We will assume this is the column
density uncertainty that propagates from the UVES extraction.

Finally, besides internal errors in the fit we have to account for another
possible systematic error, namely the continuum definition. While not
a problem in the UVES data, the continuum at the damped \lya lines
might be affected by \lya forest interlopers.  To account for this, we
performed a series of fits on spectra normalized by different QSO
continua. The modified continua were artificially displaced by $1
\sigma$ from the flux error array.  The new fits led to \ion{H}{1}
column densities that differed by $\sim 0.07$ dex from the original
values. Since internal fit errors were $0.03$ dex, and since both
errors are uncorrelated, we will assume a final 0.08 dex uncertainty
in the \ion{H}{1} measurements.



\section{Abundances in the Damped \lya System at $z=0.9313$.}

We next describe our measurements of column densities and abundances
in DLA A and B. Table~\ref{tbl-1} summarizes the results of this 
section.

\subsection{\ion{H}{1} in DLA A and B}

\begin{figure}
\epsscale{.80}
\plotone{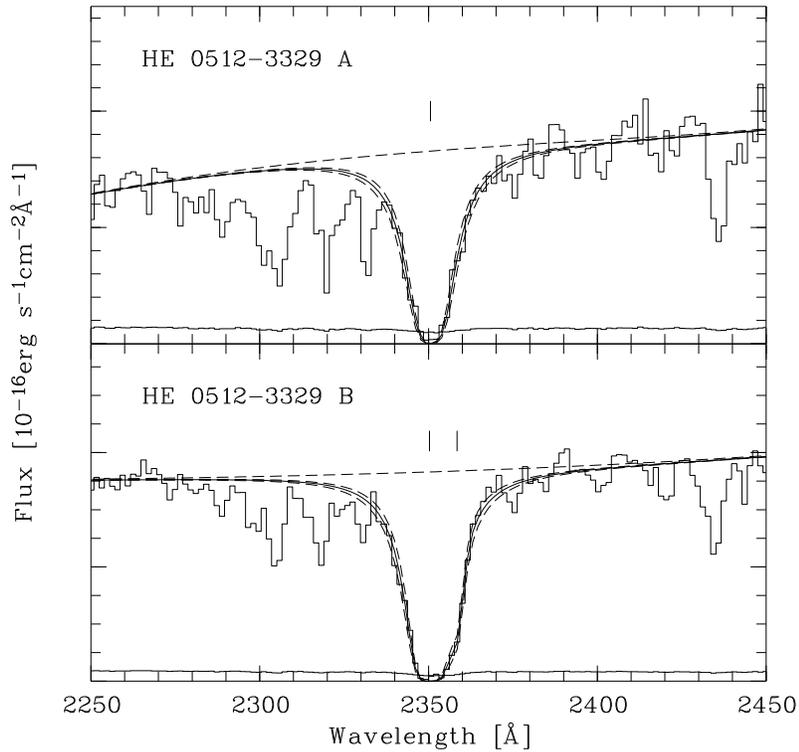}
\caption{
Portion of the HST STIS spectra of HE 0512-3329 A and B showing the
two z=0.93 damped \lya lines (histogram). The dashed horizontal line
displays the assumed QSO continuum. The smooth lines are Voigt
profiles with $\log N($\ion{H}{1}) $=20.49$ (A) and $20.47$ (B) dex,
with 0.08 dex 
deviations indicated by the dashed curves. The tick marks indicate the
fitted line centeroids. In B, an aditional HI line has been fitted
with $\log N($\ion{H}{1}) $=18.8$ dex and Doppler parameter $b=20$
\kms.  Although these values are uncertain due to the 
$b-$sensitivity on \ion{H}{1} in this column-density regime, the uncertainty
does not affect the $\log N($\ion{H}{1}) of DLA B. The separation
between the 2 tickmarks in B corresponds to 1043 km/s.
\label{fig_HI}}
\end{figure}

The \ion{H}{1} column density comes from \lya as observed in the STIS
spectra (Fig.~\ref{fig_HI}; unfortunately, the flux at the position of
\lyb\ is absorbed in either sightline by an optically thick
Lyman-limit system at $z=1.137$ -- see paper I). To obtain
$N($\ion{H}{1}) we fitted Voigt profiles to the STIS spectra. The
profiles and $1\sigma$ deviations are shown in the Figure.  Inspection
of panel B indicates blended absorption by \ion{H}{1} redwards of
$z=0.9313$, which we have included in the fit of \lya B. Fortunately,
the STIS resolution/signal-to-noise enables a fairly robust fit of the
two velocity components. Furthermore, the quite well constrained line
centroids (same results are obtained if $b$ is held fixed at different
values) are separated by a velocity difference $\Delta v = 1043 \pm
100$ \kms, which pretty much matches the $\Delta v=960$ \kms\ between
the stronger components of DLA B and a strong \ion{Mg}{2} system
observed at $z=0.9366$ in the UVES spectrum.  The fits yield $\log
N($\ion{H}{1}) $=20.49$ (A) and $20.47$ dex (B), with internal fit
errors of $0.03$ dex in both cases, and so the same values in A and B.
Significantly, a single-line fit of \lya B yields the same
$N($\ion{H}{1}) when only the blue wing of the line is used to
constrain the fit.

\subsection{Metals in DLA A and B}

Fig.~\ref{fig_lines} shows the UVES spectra of A (left panels) and B in a
velocity scale relative to $z=0.9313$.  Sightline A intercepts 7 clouds or
velocity components from $v=-129$ to $v=126$ \kms, with $\Delta v \sim 255$
\kms.  In sightline B we observe the bulk of absorption to be concentrated in
4 components around $-130$ \kms, but there are also fifth and sixth weak
components, readily seen in \ion{Mg}{2} and \ion{Fe}{2} at $+80$ and $+140$
\kms.  The total velocity span is $\approx 320$ \kms.  These kinematic
structures are similar to those ones observed in higher-redshift DLA systems.
A study of the kinematics of this system will be offered in a forthcoming
paper.  In the following, we treat each sightline separately.


\subsubsection{DLA A}

As shown in Fig.~\ref{fig_lines} the only non-saturated transitions available
in the A spectrum are \ion{Mn}{2} $\lambda 2576,2594$, \ion{Fe}{2} $\lambda
2586$, and probably \ion{Mg}{1} $\lambda 2852$.  
Below we assess the non-saturation character of the $2586$ line.  The
\ion{Mn}{2} lines are blend-free.  Integrating the $2576$ line in the range
$v=-170,150$ (this range was selected to include all \ion{Fe}{2}
components) yields an observed equivalent width of $W_{\lambda}=0.13$
\AA, and an AODM column density of $\log N($\ion{Mn}{2}$) =12.55$
\icm.  A Voigt-profile fit of the two transitions, on the other hand,
yields a total column density (3 velocity components) of $\log
N($\ion{Mn}{2}$) =12.58\pm 0.03\pm 0.04$ (internal fit error and
systematics).  We note in passing that the good fit of both transitions
reflects a minimal degree of extraction systematics.  We will adopt a
conservative value of $\log N$(\ion{Mn}{2}$)=12.58\pm 0.05$, which
error considers possible cross-talk systematics described above.
Based on this value and the \ion{H}{1} column density, the metallicity
of DLA A is [Mn/H] = $-1.44\pm0.09$ dex.

The \ion{Fe}{2} $\lambda 2586$ and \ion{Mg}{1} $\lambda 2852$ were
fitted simultaneously with 6 velocity components 
(labeled in the $\lambda 2586$ panel).
In the stronger
\ion{Fe}{2} $\lambda 2600$ transition more components are readily
detected but will not contribute significantly in the fit of the
$2586$ line.  We find $\log N($\ion{Fe}{2}$)=14.47\pm0.06\pm0.04$.
Thus, for DLA A we get [Fe/H] = $-1.52\pm0.11$ dex, and [Mn/Fe]
$=0.08\pm0.07$

Given the importance of the $\lambda 2586$ transition for the present
study, we must assess possible saturation in this line.  As pointed
out by Savage \& Sembach (1991), hidden saturation of line components can be
detected by comparing the apparent column density per unit velocity,
$N_a(v)$, of transitions of the same species but with different
$f\lambda$, where $\lambda$ is the rest-frame wavelength. In
particular, we are interested in finding out to which level the two
main velocity components in $\lambda 2586$ (2 and 3) might be
saturated. To this end,  the left-hand panel of  Fig.~\ref{fig_AODM}
shows a  comparison of the  
 iron transitions available. As expected, the $2600$ transition
(with higher $f\lambda$) yields systematically lower pixel column
densities as a consequence of it being saturated. 
If we focus on components 1 and 6 -- resolved and not
saturated {\it in 2586} --, we note that the $2600$ line reveals no
saturation in 1, while traces of saturation are evident in
6. Therefore, for the present UVES data this is roughly the optical
depth regime, $\tau\sim 1.7$, where hidden saturation becomes
important. Since component 2 in $2586$ peaks at this $\tau$, this line
should be considered slightly saturated, and the integrated AODM
column density in consequence slightly underestimated. 
 An independent test of the small saturation effect in $\lambda 2586$ is
  made in the 
  right-hand panel of Fig.~\ref{fig_AODM}, where we overplot a scaled $N_a(v)$
  obtained from the weak \ion{Mn}{2} $\lambda 2576$ line. Under the assumption of a
  nearly 
constant \ion{Mn}{2}/\ion{Fe}{2} ratio over this velocity interval, we expect
both profiles to agree within uncertainties if the $\lambda 2586$ is only
slightly saturated. It can be seen that  both profiles do show an
overall similar shape, at least to an extent where the information on $N_a(v)$
is not lost in the noise of the $\lambda 2576$ line. 
As a
consequence, we feel confident that the effect of saturation only marginally affects
the $N_{\rm AODM}$ value of \ion{Fe}{2} in Table~\ref{tbl-1}, and that
it might have completely been corrected in the adopted $N_{\rm fit}$.

\subsubsection{DLA B}

\begin{figure}
\epsscale{.80}
\plotone{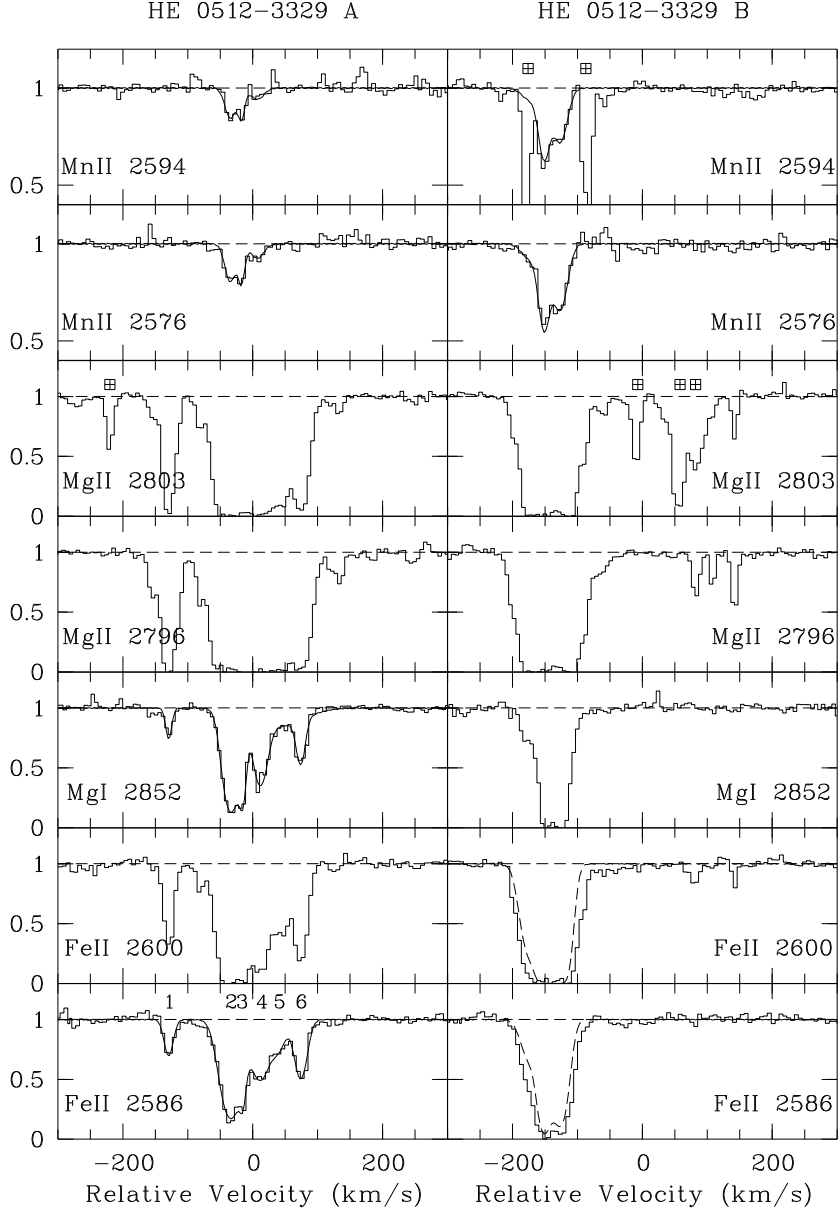}
\caption{
Absorption lines of the damped \lya system observed in the UVES spectra of
\he\ A (left panels) 
and B. The resolution is FWHM $=9.7$ \kms\ and the zero-point velocity
corresponds to $z=0.9313$. Overlaid  
are best-fit Voigt profiles whose parameters are listed in
Table~\ref{tbl-1}. Only lines believed to be non-saturated were fitted and the 
respective velocity components are labeled in the bottom-left panel.    
The dashed profiles in the two \ion{Fe}{2} lines of B were generated
with $z$ and  $b$ from the \ion{Mn}{2}(B) fits and a total column density
$\log N($\ion{Fe}{2}(B)) = $14.45$ dex, which produces [Fe/H](B) =
[Fe/H](A). Unrelated absorption features are  marked with a crossed
  square. 
\label{fig_lines}}
\end{figure}

\begin{figure}
\epsscale{1.00}
\plotone{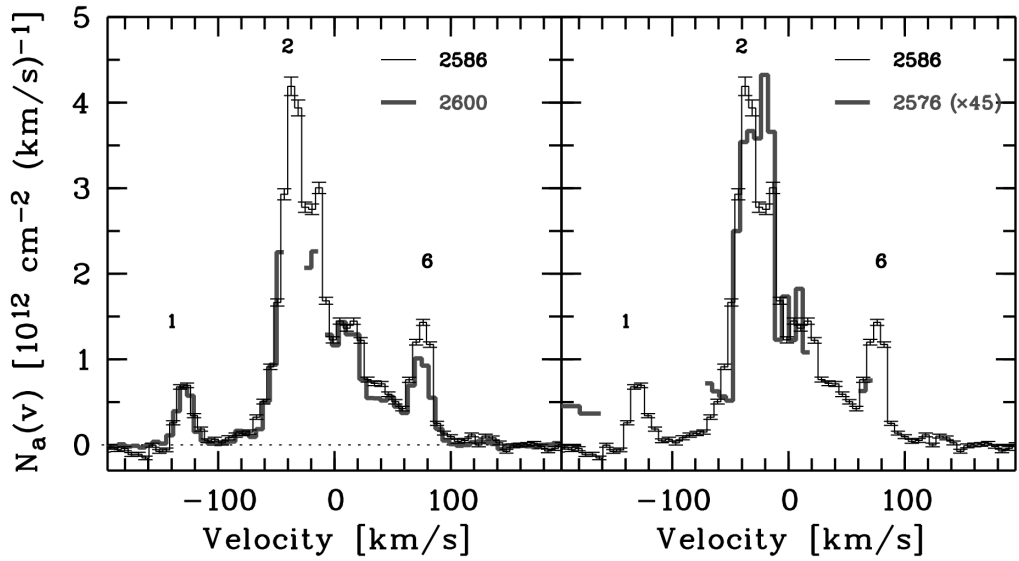}
\caption{
Apparent column density per unit velocity, $N_a(v)$, for the \ion{Fe}{2}
  $\lambda 2586$ transition  (thin histograms) and  for the \ion{Fe}{2}
$\lambda 2600$ and 
the \ion{Mn}{2} $\lambda 2576$ transitions (thick histograms) in \he\ A.
The $\lambda 2576$ profile has been scaled up by a factor of $45$. 
In both panels, $N_a(v)$ was plotted only if the normalized flux of that
pixel, $f_i$, fulfilled the condition $2\sigma_i < f_i < 1-2\sigma_i$, where 
$\sigma_i$ is the error in $f_i$. Velocity components referred to in the text
are labeled. 
\label{fig_AODM}}
\end{figure}

As shown in Fig.~\ref{fig_lines} the only non-saturated transitions
available in the B spectrum are \ion{Mn}{2} $\lambda 2576,2594$. The
$2594$ line is blended with the $z=1.13$ \ion{Fe}{2} $\lambda 2344$
line. Integrating the $2576$ line in the range $v=-210,-80$ yields an
equivalent width of $W_{\lambda}=0.33$ \AA, and an AODM column density
of $\log N($\ion{Mn}{2}$) =12.99$ \icm. A Voigt-profile fit of the two
transitions, on the other hand, yields a total column density (3
velocity components) of $\log N($\ion{Mn}{2}) $=13.02\pm 0.01\pm
0.04$. Note that the synthetic profiles show readily slight
discrepancies, which we believe are not due to innacurate atomic data
but instead can be attributed to cross-talk extraction problems. A fit
of \ion{Mn}{2} $\lambda 2576$ only yields $\log N($\ion{Mn}{2}$)
=13.00$, which is within our estimate of those systematic errors. We
will conservatively adopt $\log N($\ion{Mn}{2}$) =13.02$. Based on this
value, the metallicity of DLA B is therefore [Mn/H] = $-0.98\pm0.09$
dex.

The two \ion{Fe}{2} transitions in the UVES spectrum are saturated; therefore,
from the integrated apparent optical depth we only get a quite conservative
lower limit of $\log N$(\ion{Fe}{2}) $>14.65$, or [Fe/H] $>-1.32$ dex, and
[Mn/Fe] $<0.34$ dex.  A Gaussian fit to the much weaker ( though saturated )
$\lambda 2374$ line in the STIS spectrum yields an observed equivalent width
of $W_\lambda=1.33\pm 0.30$ \AA, which translates into the same lower limit
for $N($\ion{Fe}{2}).  In a different approach to estimate the Fe abundance,
we generated synthetic profiles of these two \ion{Fe}{2} lines by using the
same number of components, $b$-values and redshifts obtained from the
Voigt-profile fits to the \ion{Mn}{2} transitions. \ion{Fe}{2} Doppler
parameters in excess of those of \ion{Mn}{2} are not physical.  The
\ion{Fe}{2} column density was chosen to get [Fe/H](B) = [Fe/H](A) = $-1.52$.
The resultant profiles are superimposed with dotted lines in the two
\ion{Fe}{2} B panels of Fig.~\ref{fig_lines}.  Comparing with the data, the
absorption excess is evident; moreover, it cannot be accounted for by
increasing $b$ even by several $\sigma_b$'s, nor can it be the result of an
extraction artifact. The Fe column density of the B profiles has to be further
increased by $\approx 0.4$--$0.5$ dex to match the data ([Fe/H] $\approx
-1.1$-- $-1.0$)\footnote{Alternatively, additional weak velocity components
---detected in \ion{Fe}{2} but not in \ion{Mn}{2}--- might also contribute to
the extra absorption.}. This analysis indicates that, first, [Fe/H](B) $>$
[Fe/H](A) without any doubt; and second, the excess metallicity in B {\it
might} be in line with that for \ion{Mn}{2}, so that [Mn/Fe](A) $\approx$
[Mn/Fe](B).

\section{Discussion}

\subsection{Metallicity at $z=0.93$ and the DLA galaxy}

There are currently only 5 measurements of metallicity in DLAs at
$z<1$ (Prochaska et al.\ 2003a). The $z=0.9313$ DLA in \he\ adds an
important value to the overall sample at an epoch for which models of
chemical evolution can distinguish different kinds of galaxy
morphologies (Calura, Matteucci \& Vladilo 2003).  The metallicity
derived for DLA A, [Fe/H] $=-1.52$, is representative of the $z\approx
1 $ sample, and the data on Mn and Fe abundances either in A or B are
consistent with a low-metallicity and dusty absorber, as we shall
discuss below.

Because the galaxy hosting the $z=0.9313$ absorbers is also the likely lensing
galaxy, DLA A and B then probe the ISM on two opposite sides and the impact
parameters cannot be larger than $2.5$ \hkpc.  These small distances are
counterexamples to the claim that lines-of-sight close to DLA galaxies are
missing from DLA samples due to dust obscuration (see discussion in Ellison et
al.\ 2001).  A further example is the $z=0.807$ QSO HS 1543$+$5921, which
line-of-sight intersects the DLA galaxy ---a low surface brightness galaxy at
$z=0.009$--- at only $0.5$ \hkpc\ (Reimers \& Hagen 1998; Bowen, Tripp, \&
Jenkins 2001), yet without any detectable reddening.

Deep imaging of DLA
fields has uncovered an inhomogeneous population of gas-rich galaxies with a
range of morphologies and luminosities (e.g., Le Brun et al.\ 1997). For \he,
Gregg et al.\ estimated the lens to be roughly consistent with an $L^{*}$
galaxy.  It is remarkable to find such a low-metallicity absorber so close to a
luminous galaxy.  Discrepancies between gas-phase and emission-line abundances
have recently been studied by Ellison et al.\ (2004b).


\subsection{Two Damped \lya Systems, One Reshift}

The real value of \he\ as a DLA probe comes from its double nature.  Although
perhaps a rare situation, the dual sight-line abundances obtained here allow
us to reach some important conclusions.

The present data on \he\ are consistent with transverse differences in
{\it both} [Mn/H] and [Fe/H] at $z=0.9313$.  For Mn, this gradient
amounts $0.46\pm0.13$ dex (fit and systematic errors) on a spatial
scale of $\sim 5$ \hkpc\, and with sightline A passing though the more
metal-deficient gas-phase\footnote{We have noted that both sightlines
cross other absorbing clouds at several $100$ \kms. In particular, the
A spectrum shows a weak \ion{Mg}{2}-\ion{Fe}{2} system that is separated by
$400$ \kms\ from the DLA. Although such a large velocity span likely
implies two separated objects, even inclusion of this system would
have minimal impact ($<5$ \% in column density) on the abundances
derived for DLA A.  Thus, the [Mn/H] gradient cannot be explained by a
particular choice of velocity components.}.  Our analysis shows that a
similar gradient is present also in [Fe/H], although the observational
uncertainties are larger.

Interestingly, [Mn/Fe](A) $\approx$ [Mn/Fe](B), which, if not greatly
affected by dust ---see below---, suggests chemical uniformity across
the lines of sight at much larger distances than probed before
(Churchill et al.\ 2003; Ellison et al.\ 2004a) and in accordance with
uniformity along the lines of sight (Lopez et al.\ 2002; Prochaska
2003b).

\subsection{Dust}

Abundance patterns observed in DLAs are a consequence of both
nucleosynthetic yields and differential incorporation of atoms into
dust grains in the DLA ISM.  Being an Fe-peak element, Manganese
shares a common origin with Fe; however, a manganese underabundance is
expected from the nucleosynthesis odd-even effect (Truran \& Arnett
1971), borne out by measurements of Galactic disk and halo star
abundances (Lu et al.\ 1996 and references).  On the other hand,
[Mn/Fe] $>0$ in diffuse ISM clouds of the Galactic disk for all range
of metallicities (Savage \& Sembach 1996); this is because iron is
more prone than Mn to dust depletion, and consequently underabundant
in the gas phase.
 
Our analysis of DLA A indicates [Mn/Fe] $=0.08\pm0.07$.  This ratio,
consistent with solar, is among the highest of all [Mn/Fe] ratios at every
redshift (Ledoux et al. 2002).  Ledoux et al.  have investigated in detail
dust depletion effects on Mn; in particular they have noted a correlation of
[Mn/Fe] with metallicity, which they interpret as the link between
nucleosynthesis yields in DLAs and abundances in halo stars.  In this regard
(see their Figure 4) DLA A is a clear outlier, with manganese being far too
abundant for [Fe/H] $=-1.5$.  The only explanation for this high value is
significant dust depletion in DLA A.  Following the their trend of [Mn/Fe] with
[Zn/Fe] (the dust-to-gas ratio) iron in DLA A might be depleted by as much as
0.8 dex.

We have not been able to accurately quantify [Mn/Fe] in DLA B, but the
present data are consistent with a similar ratio as in DLA A.  This
solar ratio {\it might} be taken as evidence for dust also in B since
all observed DLAs have underabundant Mn unless [Zn/Fe] $>\sim 0.6$, an
established indicator of dust depletion.  
However, the
cosmic scatter of [Mn/Fe] at [Fe/H] $=-1$ is larger and there are some
examples of high [Mn/Fe]  even at low [Zn/Fe].  Furthermore, 
our derived lower limit on Fe/H translates into a potentially lower   Mn/Fe
ratio, more in line with the general trend of DLAs at the low-depletion end.

\subsection{Gradient in Metallicity?}

Supposing that the [Mn/Fe] in both A and B implies homogeneously
distributed dust, a real metallicity gradient appears at a first
glance as an attractive explanation for the transverse differences in
[M/H]. In nearby spirals metallicity gradients as a function of
galactocentric radius are found in the range $-0.04$ to $-0.20$ dex
kpc$^{-1}$ (e.g., Vila-Costas \& Edmunds 1992).  That the two images
of \he\ are of similar brightness and that the lens is probably the
DLA galaxy under study both imply the absorbing regions must lie more
or less equidistant from the galaxy center, not farther than $2.5$
\hkpc  (the lensing galaxy has so far not been detected, so its exact
location is undetermined).  This is consistent with the similar
\ion{H}{1} in DLA A and B.  A simple model of varying metallicity with
radius, therefore, cannot explain the differences observed at similar
impact parameters.

\subsection{Differential Dust Depletion/Reddening?}

Aternatively, differential dust depletion offers a plausible
explanation of the abundance differences.  As stated above, in A there
is dust depletion by a factor of $\sim 3$.  We now suppose that, on the
contrary, in B the depletion is minimal, which is still consistent
with [Mn/Fe] $\approx 0$ given [M/H] $\approx -1$.  There is enough
evidence that at least some DLAs are bound to rotating systems,
so some sightlines should encounter DLAs in disk+halo
configurations.  One such examples have been found by Quast, Reimers \&
Baade (2004) in the $z=1.15$ DLA toward HE 0515$-$4414.  The velocity
components there cover the whole range of depletion patterns from Warm
Halo ([Cr/Zn] $=-0.5$ dex) to Warm Disk ($-1.1$) observed in our
galaxy.  In \he, the abundance pattern of DLA A closely resembles that
of the the Galactic Warm Halo (see Fig.6 in Savage \& Sembach), with
Mn and Fe depleted by a factor of 3--4.  In consequence, the different
Fe and Mn abundances can be explained by lines of sight crossing Warm
Halo-like regions that are subject to distinct dust depletion factors,
with DLA B being less dusty.

In this scenario, the observed differences of [M/H] in \he, where M
are refractory elements, become a different and direct confirmation
for the hypothesis of dust-depletion in DLAs.  Moreover, we would
expect differential reddening between QSO A and B (Pei, Fall, \&
Bechtold 1991).  Photometric data of \he\ (Gregg et al.) in fact show
that extinction along the line of sight to QSO A is greater, although
as pointed out in Paper I, a chromatic effect of
microlensing could also be present.  As a rough estimate, $\log
N($\ion{H}{1}$)=20.5$ dex leads to an expected color-excess of
$E(B-V)=0.06$ if dust is present. With an SMC type extinction curve we
get $E(1250-V)/E(B-V)=14$ (this is the shortest reliable wavelength of
the extinction curve), which gives $A(\lambda_{\rm obs}=2400 {\rm
\AA})=0.84$ mag for an extinction factor of 2.2.  This agrees with the
$f_{\rm A}/f_{\rm B}$ ratio from the STIS spectra plus an aditional
contribution due to microlensing (Paper I).
The underlying assumption here is that the dust responsible for the
depletion of Fe in DLA A is the same as for the extinction, a reasonable
approach since extinction in the UV is dominated by small dust grains like
iron oxides (see Prochaska \& Wolfe 2002 for a complete discussion). 

Finally, one concern on the differential-dust hypothesis: 
high-redshift DLAs in general do not appear
to produce significant dust-redenning (Murphy \& Liske 2004). In consequence, 
either dust at $z=0.93$ is not the main agent of the differential
reddening observed in \he, or DLA A is not representative of the systems
studied by Murphy \& Liske. 


\subsection{Final Remark}

Regardless of the extent to which dust-depletion is acting
differentially to alter [Fe/H], one key consequence of the sightline
differences in \he\ is their connection to the cosmic spread in
[Fe/H].  The present sample of metallicity measurements at all
redshifts is distributed within [Fe/H] $= 0,-3$ dex (see Prochaska et
al.\ 2003a for the latest compilation).  The overall spread is
a consequence of a mild but readily noticeable evolution of metals over
a Hubble time. Petinni et al.\ (1994), however,
found that even sub-samples within narrow redshift bins show large scatter of
more than 1 dex.  This is not attributable only to dust-depletion ([Zn/H]
vs.\ $z$ shows a similar behaviour, and zinc is not depleted);
different star formation histories must contribute to the
inhomogeneity.  Despite this, if the DLA in front of \he\ is
representative of $z\approx1$ DLAs, 
then the
present sample of DLA abundances must be randomly biased by effects other
than galaxy type, such as the specific evolutionary stage in
a particular line of sight, perhaps a high recent SN rate, or high
star-formation rate with many UV emitting stars, which in the case of DLA
B may have inhibited dust formation.  
In the overall context this would mean that the "patchiness" of star
formation in a given galaxy dominates which type of DLA we observe;
furthermore, this stochastic ingredient could be more marked at higher
redshifts, where mixing has had less time to occur.  In conclusion,
part of the cosmic scatter must be due to these local effects which
cannot be studied comprehensively.
Obtaining better-quality optical spectra of \he\ A and B, especially
in the spectral region covering \ion{Zn}{2} and \ion{Cr}{2}
transitions, could offer important diagnostics for our understanding
of abundances in DLAs as a whole.

\section{Summary}

We have studied the damped \lya system at $z=0.9313$ in front of QSO
\he. This system is particularly advantageous for new insights into
DLA studies because it is probed by two lines of sight, most likely on
opposite sides of the lensing galaxy.  We have confirmed the damped
nature of the system in the two lines of sight and obtained \ion{H}{1}
column densities with help of spatially resolved STIS spectroscopy.
Then, in a second result, we have obtained column densities of
\ion{Fe}{2} and \ion{Mn}{2} in both lines of sight from ground-based
echelle spectroscopy.  This DLA becomes only the sixth case at $z<1$
with measured metallicity.  We have assessed and quantified the
sources of systematic error in the column densities, given (1) the
challenging character of the data reduction and (2) the small set of
transitions that could be measured.  We inspected in detail cross-talk
effects in the echelle data, continuum estimates in the STIS data, and
saturation effects in the reduced spectra.  Armed in the end with a
small set of Mn and Fe abundances in each of the sightlines, our third
result is the discovery of significant differences, $\sim 0.5$ dex in
both [Mn/H] and [Fe/H], on a transverse scale of $5$ \hkpc, while
[Mn/Fe] appears solar in both A and B.  Our treatment of errors leaves
no doubt that the gradient is real, and we have discussed two possible
causes: different nucleosynthesis yields and differential
dust-depletion.  Regardless of what produces the gradient in this
particular DLA, our finding offers a compelling explanation for the
source of scatter in the observed evolution of metals in the neutral
Universe.

\acknowledgments

We have benefitted from comments made by Sara Ellison, Cedric Ledoux, and
Celine Peroux. We are also grateful to the anonymous referee for a very useful
report.  SL would like to also thank the ESO Scientific Visitor Program for
supporting a pleasent stay at ESO Headquarters, where part of this work was
done.  SL also acknowledges support from the Chilean {\sl Centro de Astrof\'\i
sica} FONDAP No. 15010003, and from FONDECYT grant N$^{\rm o} 1030491$. OW is
funded by the BMBF/DLR Verbundforschung under grant 50~OR~0208.  Additional
funding for this work was provided by NASA through grant number GO-9165 from
the Space Telescope Science Institute, which is operated by AURA, Inc., under
NASA contract NAS5-26555.

\clearpage

\clearpage

\clearpage

\clearpage

\clearpage

\clearpage

\begin{deluxetable}{lcccccccccc}
\tabletypesize{\scriptsize}
\tablecaption{Line Parameters in the $z=0.9313$ Damped \lya System  Toward \he \label{tbl-1}}
\tablewidth{0pt}
\tablehead{
&&&\multicolumn{4}{c}{\he\ A}&\multicolumn{4}{c}{\he\ B}\\
\colhead{Ion}&\colhead{$\lambda_0$}&\colhead{$f$}&\colhead{$W_{\lambda}$}
& \colhead{$\log N_{\rm AODM}$} &
\colhead{$\log N_{\rm fit}$} & \colhead{[X/H]} &
\colhead{$W_{\lambda}$} & \colhead{$\log N_{\rm AODM}$} &
\colhead{$\log N_{\rm fit}$} &  \colhead{[X/H]}
}
\startdata
\ion{H}{1} & 1215.6701 & 0.4164 & 12.62 & ...            &20.49$\pm$0.08 & ... & 12.60& ...&20.47$\pm0.08$  & ...   \\
\ion{Mg}{1} & 2852.9642 & 1.8100 & 1.37 &
12.94$\pm$0.01&13.01$\pm$0.02 & ... & 1.14 & $>$12.95 & ... & ...   \\
\ion{Mg}{2} & 2796.3520 & 0.6123 & 3.75 &  $>$14.11   & ... & ... & 2.03 & $>$13.90 & ... & ...   \\
\ion{Mn}{2} & 2576.8770 & 0.3508 & 0.13 &12.55$\pm$0.01&12.58$\pm$0.03
&-1.44$\pm$0.09& 0.33 &12.99$\pm$0.01 &13.02$\pm$0.01 & -0.98$\pm$0.09
\\ 
         & 2594.4990 & 0.2710 &0.09 &&&&...&&\\ 
\ion{Fe}{2} & 2586.6500 & 0.0691 & 1.33 &14.41$\pm$0.01 &14.47$\pm$0.06 &-1.52$\pm$0.11 & 1.23 &$>$14.65 &... & $>$-1.32
\\ 

\enddata


\tablecomments{Column density errors are internal fit errors only;
  errors in [M/H] include fit and  systematic errors.
}

\end{deluxetable}



\begin{thebibliography}{}

\bibitem[]{ba94}
Ballester, P. 1994 A\&A, 286, 1011

\bibitem[]{boi98}
	Boiss\'e, P., Le Brun, V., Bergeron, J., \& Deharveng, J.-M.,
	1998, A\&A, 333, 841

\bibitem[]{bo01}
Bowen, D. V., Tripp, T. M., \&  Jenkins, E. B., 2001, AJ, 121, 1456

\bibitem[]{ca03}
Calura, F., Matteucci, F., \& Vladilo G. 2003, MNRAS, 340, 59 

\bibitem[]{cwc03}
        Churchill, C. W., Mellon, R. R., Charlton, J. C., \& Vogt, S.,
 	2003, ApJ, 593, 203


\bibitem[]{el01} Ellison, S. L., Yan, L., Hook, I. M., Pettini, M., Wall,
  J. V., \& Shaver, P. 2001, \aap, 379, 393 

\bibitem[]{el04a} Ellison, S. L., Ibata, R., Pettini, M., Lewis, G. F.,
Aracil, B., Petitjean, P., \& Srianand, R. 2004a, A\&A, 414, 79

\bibitem[]{el04b} Ellison, S. L., Kewley, L. J., \& Mall\'en-Ornelas,
  G. 2004b, submitted 

\bibitem[]{fo95}
Fontana, A., \& Ballester, P. 1995, The Messenger, 80, 37

\bibitem[]{gr00}
Gregg, M. D., Wisotzki, L., Becker, R. H., Maza, J., Schechter, P. L., White,
R. L., Brotherton, M. S., \& Winn, J. N. 2000, AJ, 119, 2535

\bibitem[]{gs98}
        Grevesse, N., \& Sauval, A.J. 1998, Space Sci Rev, 85,
        161

\bibitem[]{kob02}
        Kobayashi, N., Terada, H., Goto, M., Tokunaga, A., 2002, ApJ,
	569, 676


\bibitem[]{le97}Le Brun, V., Bergeron, J., Boiss\'e, P., \& Deharveng, J. M.,
  1997, A\&A, 321, 733 

\bibitem[]{le00}Le Brun, V., Smette, A., Surdej, J., \& Claeskens, J.-F.,
  2000, A\&A, 363, 837 

\bibitem[]{lbp02}
	Ledoux, C., Bergeron J., \& Petitjean, P., 2002,
	A\&A, 385, 802

\bibitem[]{lo99}
Lopez, S., Reimers, D., Rauch, M., Sargent, W.L.W., \& Smette, A.
1999, \apj, 513, 598

\bibitem[]{lo02}
Lopez, S., Reimers, D., D'Odorico, S., \& Prochaska, J.X. 2002, 
\aap, 385, 778

\bibitem[]{lu96}Lu, L., Sargent, W. L. W., Barlow, T. A., Churchill, C. W., \&
  Vogt, S. 1996, ApJS, 107, 475 

\bibitem[]{mo69}
Moffat, A. F. J. 1969, A\&A, 3, 455 

\bibitem[]{mo00}
Morton, D. C.  2000, ApJS, 130, 403 

\bibitem[]{ml04}
Murphy, M. T., \& Liske J.  2004, MNRAS, 354, 31


\bibitem[]{pei91}  
Pei, Y.C., Fall, S.M., \& Bechtold, J. 1991, \apj 378, 6


\bibitem[]{ptt94} 
Pettini, M., Smith, L. J., Hunstead, R. W., and King,
  D. L. 1994, \apj, 426, 79

\bibitem[]{889} Prochaska, J. X., \& Wolfe, A. M. 1998, ApJ, 507, 113

\bibitem[]{pw02}
Prochaska, J.X. \& Wolfe, A.M. 2002, \apj, 566, 68


\bibitem[]{895} Prochaska, J. X., Gawiser, E., Wolfe, A. M., Castro, S., \& 
Djorgovski, S. G. 2003a, ApJ, 595, L9

\bibitem[]{pro03a}  	
Prochaska, J.X. 2003b, \apj, 582, 49

\bibitem[]{qu04}
Quast, R., Reimers, D., \& Baade, R. 2004, in preparation


\bibitem[] {}
Rao, S., Nestor, D., Turnshek, D.,  Lane, W., Monier, E.,
\& Bergeron, J. 2003, ApJ, 595, 94

\bibitem[]{912}
Rao, S., Nestor, D., Turnshek, D.,  Lane, W., Monier, E.,
\& Bergeron, J. 2003, ApJ, 595, 94



\bibitem[] {}
Reimers, D. \& Hagen, H.-J. 1998, A\&A 329, L25 


\bibitem[]{916}
Savage, B. D., \& Sembach, K. R. 1991, ApJ, 379, 245

\bibitem[]{919}
Savage, B. D., \& Sembach, K. R. 1996, ARA\&A, 34, 279

\bibitem[]{sm95}
        Smette, A., Robertson, J. G., Shaver, P., Reimers, D., 
	Wisotzki, L., \& Kohler, Th., 1995, A\&AS, 113, 199

\bibitem[]{sm97}
        Smette, A., Claeskens, J.-F., \& Surdej, J.
	1997, NewA, 2, 53

\bibitem[]{ta71}
Truran, J. W. \& Arnett, W. D. 1971, Ap\& SS, 11, 430

\bibitem[]{vi92}
Vila-Costas, M. B.,  \& Edmunds, M.G. 1992, MNRAS, 259, 121

\bibitem[]{wu03}
Wucknitz, O., Wisotzki, L., Lopez, S., \& Gregg, M. D. 2003, A\&A 405, 445
(Paper I)

\bibitem[]{zuo97}
        Zuo, L., Beaver, E. A., Burbidge, E. M., Cohen, R., D., Junkkarinen, 
	Vesa, T., Lyons, R. W., 1997, ApJ, 477, 568 


\end{thebibliography}
\end{document}